\documentclass[a4paper]{article}
\usepackage{amsfonts}
\usepackage{amssymb}
\usepackage{amsmath}
\usepackage[dvips]{graphics}
\usepackage{latexsym}
\usepackage{geometry}
\usepackage{verbatim}
\usepackage{authblk}

\newcommand{\bean}{\begin{eqnarray}}
\newcommand{\eean}{\end{eqnarray}}

\newcommand{\eq}[1]{Eq. (\ref{#1})}

\newcommand{\pp}[2]{\frac{\partial #1}{\partial #2}}

\newcommand{\bea}{\begin{eqnarray*}}
\newcommand{\eea}{\end{eqnarray*}}
\newcommand{\grad}{\nabla}

\newcommand{\eqn}{&=&}
\newcommand{\non}{\nonumber \\}

\begin{document}
\title{General Proof of the Tolman law}
\author{Minghao Xia}
\author{Sijie Gao\thanks{Corresponding author: sijie@bnu.edu.cn}}
\affil[1]{Department of Physics, Beijing Normal University, Beijing , China,100875}

\maketitle

\newtheorem{theo}{Theorem}

\begin{abstract}
Tolman proposed that the proper temper $T$  of a static self-gravitating fluid in thermodynamic equilibrium satisfies the relation $\chi T=constant$, where $\chi$ is the redshift factor of the spacetime. The Tolman law has been proven for radiation in stationary spacetimes and for perfect fluids in stationary, asymototically flat and axisymmetric spacetimes. It is unclear whether the proof can be extended to more general cases. In this paper, we prove that under some reasonable conditions, the Tolman law always holds for a perfect fluid in a stationary  spacetime. The key assumption in our proof is that the particle number density $n$ can not be determined by the energy density $\rho$ and pressure $p$ via the equations of state. This is true for many known fluids with the equation of state $p=p(\rho)$. Then, by requiring that the total entropy of the fluid is an extremum for the variation of $n$ with a fixed metric, we prove the Tolman law. In our proof, only the conservations of stress energy and the total particle number are used, and no field equations are involved. Our work suggests that the Tolman law holds for a generic perfect fluid in a stationary spacetime, even beyond general relativity.

\end{abstract}

\section{Introduction}
 In 1930, Tolman \cite{Tolman} investigated the thermodynamic implications of the mass-energy  relation proposed by Einstein and found that the local temperature $T$ satisfies $ T\chi=constant$, which is called Tolman law, where $\chi$ is the redshift factor. Later,  Tolman and Ehrenfest proved that it also holds in a general static gravitational field\cite{TE}. However, this proof is essentially for radiation, not for a general perfect fluid. In 1949,  Buchdahl further extended this conclusion to a stationary gravitational field, but only for the case of radiation\cite{Buchdahl}. In 2019, Lima et al. derived Tolman law in a static spacetime without specific equations of state \cite{Lima}. Again, the proof holds for radiation. For nonvanishing chemical potential $\mu$, some extra assumptions are needed.

It is also worth mentioning  some different approachs to derive the Tolman law. Ebert and Gobel obtained the Tolman law through the extension of the thermal law and Carnot cycle\cite{Ebert}. Rovelli and Smerlak showed that the Tolman law follows from applying the equivalence principle to the thermal time\cite{Rovelli}.

So far, the most general and rigorous proof of the Tolman law was given by Green, Schiffrin and Wald \cite{wald13}. They showed that in a stationary, axisymmetric, asymptotically flat spacetime, the temperature distribution of a perfect fluid satisfies the Tolman law. The proof is obtained by requiring that the total entropy of the fluid is an extremum for fixed total energy and angular momentum. The asymptotic flatness of the spacetime guarantees a well-defined total energy or angular momentum. It is then natural to ask whether we can have a local version of the proof without involving global properties of the spacetime? This is not easy since it is well known that energy cannot be localized in general relativity. In this paper, we shall bypass the definition of energy. We focus on any finite region $\Sigma$ which is filled up with stationary perfect fluid. We wish to obtain the Tolman law by extremizing the total entropy of fluid in this region. Usually, the extremization of entropy needs constraints on energy and angular momentum. To find alternative constraints, we notice  the fact that, for most fluids, the particle number density and temperature are intertwined in the equations of states. Moreover, only the energy density $\rho$ and pressure $p$ appear in the stress-energy tensor, which is related to the spacetime metric by the field equation. Therefore, given the spacetime metric in $\Sigma$, only the distributions of $\rho$ and $p$ are fixed and $n$ can be freely specified without changing the spacetime. Then, the best way to single out a distribution $n$ is to maximize the total entropy. We show, as in the following theorem, that this just leads to the Tolman law.

\section{Proof of the Tolman law}
Now we express the Tolman law in the following theorem.

\begin{theo}
  Assume that the spacetime $(M,g_{ab})$ is stationary, i.e., there exists a timelike Killing vector field $\xi^a$ in a spacetime region $\Sigma$. Let $\chi=\sqrt{-\xi^a\xi_a}$ be the redshift factor. Then we have $\xi^a=\chi u^a$, where $u^a$ is the 4-velocity of the stationary observers. Suppose that the matter in the spacetime is a perfect fluid with the stress-energy tensor
\bean
T_{ab}=\rho u_au_b+p(g_{ab}+u_au_b)\,,
\eean
where $\rho$ is the energy density and $p$ is the pressure of the fluid. Then the Tolman law $\chi T=constant$ holds in $\Sigma$  if the following conditions are satisfied:

\noindent 1. $\rho$ and $p$ are determined by the spacetime metric $g_{ab}$ via the field equation.

\noindent 2. Given the distributions of $\rho$ and $p$ in $\Sigma$, the particle number density $n$ and the temperature $T$ cannot be solved from the equations of state.

\noindent 3. The total entropy in $\Sigma$ is an extremum for all configurations of the matter with fixed metric and total particle number $N$.
\end{theo}

Before we prove the theorem, we would argue that all the three conditions are natural. Condition 1 is obviously true for general relativity and many modified theories of gravity. To understand Condition 2, we remember that the equations of state for Fermi and Bose gases are \cite{landau}
\bean
p\eqn nT\left(1\pm\frac{\pi^{3/2}}{2g}\frac{n\hbar^3}{(mT)^{3/2}}\right) \,,\\
p\eqn\frac{2}{3}\rho \,.
\eean
Obviously, the distributions of $\rho$ and $p$ cannot determine the distributions of $n$ and $T$. We believe that most perfect fluids with the equation of state $p=p(\rho)$ satisfy condition 2. A counterexample is the radiation. However, we shall show below that the Tolman law is automatically satisfied for radiation in a stationary spacetime. If these two conditions hold, we see immediately that the spacetime metric cannot give the distribution of $n$ or $T$. Then it is natural to require that the distribution of $n$ or $T$ maximize the total entropy for a fixed spacetime metric, which is just Condition 3.

Now we prove the theorem. From the conservation of stress energy $\grad_aT^{ab}=0$, one can show that for stationary fluid \cite{gao14}
\bean
\grad_a p=-(\rho+p)\grad_a\chi/\chi \label{gradp}\,.
\eean
In fact, for radiation, $p=\rho/3$ and $\rho=bT^4$, \eq{gradp} implies immediately the Tolman law $T\chi=const$.

The total entropy is given by
\bean
S=\int_\Sigma \sqrt{h} s(x^i) d^3 x\,,
\eean
where $h$ is the determinant of the spatial metric $h_{ab}=g_{ab}+u_au_b$ and $\{x^i\}$ represent the spatial coordinates.
The total number of particle is
\bean
N=\int_\Sigma \sqrt{h} n(x^i) d^3 x\,.
\eean
Since the total number of particle is fixed, an extremum of the total entropy gives \cite{gao11}
\bean
\delta S+\lambda \delta N=0\,,
\eean
which can be rewritten as
\bean
\delta \int_\Sigma L=0\,.
\eean
Since $\rho$ depends on the metric and its derivatives, we can assume
\bean
L(\rho, n, g_{\mu\nu})=s(\rho, n)\sqrt{h}+\lambda n\sqrt{h}\,,
\eean
where we have fixed the vector field $\xi^a$ and then $h_{ab}$ is determined by $g_{ab}$.
Note that by the field equation, $\rho$ can be expressed as a function of the metric. Thus $L$ can be written as
\bean
L=L(n,g_{\mu\nu}) \,.\label{llng}
\eean
According to our assumption, the metric is fixed and $n$ is independent of the metric. Therefore, the extremum of the total entropy in $\Sigma$ leads to
\bean
\pp{L}{n}=0\,,
\eean
which means
\bean
\pp{s}{n}+\lambda=0\,.
\eean
From the first law \cite{gao11}
\bean
ds=\frac{1}{T}d\rho-\frac{\mu}{T}dn \label{firstlaw}
\eean
we find
\bean
-\frac{\mu}{T}+\lambda =0 \label{mut}\,.
\eean

From the Gibbs-Duham relation
\bean
p=Ts+\mu n-\rho\,,
\eean
we have
\bean
dp =Tds+sdT+\mu dn+nd\mu-d\rho\,.
\eean
Together with \eq{firstlaw}, we find\cite{gao11}
\bean
dp=sdT+nd\mu \label{dpst}\,.
\eean
Substitution of \eq{mut} into \eq{dpst} yields
\bean
dp\eqn sdT+n\lambda dT \non
\eqn \left(s+n\frac{\mu}{T}\right)dT \non
\eqn \frac{p+\rho}{T}dT\,,
\eean
or
\bean
\grad_a p=\frac{p+\rho}{T}\grad_a T\,.
\eean
Comparing with \eq{gradp}, we finally obtain the Tolman law
\bean
T\chi=const\,.
\eean

Following \eq{mut}, we see that the Tolman law also implies the distribution of chemical potential $\mu\chi =constant$.

\section*{Discussions}
Our proof of the Tolman law applies to most stationary perfect fluid, not relying on the field equation and the definition of total energy. Only the local conservation of energy and the conservation of total particle number are assumed. Therefore, the Tolman law could be a result of the extremization of total entropy, regardless of the specific theory of gravity. We have replaced the constraints on the total energy and angular momentum by requiring that the metric, i.e., the spacetime, is fixed. This is a natural treatment and can be understood as follows. If we have a fluid distributed in a region of spacetime and the distribution of particle number density $n$ can not be dermined by the field equation, then how can we know the distribution of $n$? Among all the configurations of $n$ with the same spacetime, the one that maximizes the total entropy obviously should be the favorite, which just leads to the Tolman law according to our theorem. In other words, if the Tolman law were not true, the total entropy would not be an extremum. The key assumption in our proof is that the equation of state takes the form $p=p(\rho)$ and then the particle number density can be specified freely for fixed distributions of $p$ and $\rho$. Although this assumption is true for most known fluids, it would be interesting to see whether the Tolman law still holds if the condition is violated. One may wonder what if the metric is not fixed. Of course, we can vary the metric to maximize the entropy as we did in \cite{gao14}, which leads to the field equation. However, we have just shown that the derivation of the Tolman law is irrelevant to the field equation.

\end{document}